\documentclass[11pt]{article}
\usepackage[utf8]{inputenc}
\usepackage[T1]{fontenc}
\usepackage{natbib}
\usepackage{booktabs}
\usepackage{rotating}
\usepackage{subfigure}
\usepackage{multirow}
\usepackage{graphicx}
\usepackage[official]{eurosym}
\usepackage{accessibility}
\usepackage{amsmath}
\usepackage{amssymb}
\usepackage{url}
\usepackage{breakurl}

\providecommand{\keywords}[1]{\textbf{Keywords:} #1}

\usepackage[a4paper, top=1in, bottom=1in, left=0.55in, right=0.55in]{geometry}

\title{Exploring User Acceptance and Concerns toward LLM-powered Conversational Agents in Immersive Extended Reality}
\author{
  Efe Bozkir$^{1}$ \and
  Enkelejda Kasneci$^{1}$
}
\date{}

\begin{document}
\maketitle

\begin{center}
$^{1}$Technical University of Munich\\
\end{center}

\begin{abstract}
The rapid development of generative artificial intelligence (AI) and large language models (LLMs), and the availability of services that make them accessible, have led the general public to begin incorporating them into everyday life. The extended reality (XR) community has also sought to integrate LLMs, particularly in the form of conversational agents, to enhance user experience and task efficiency. When interacting with such conversational agents, users may easily disclose sensitive information due to the naturalistic flow of the conversations, and combining such conversational data with fine-grained sensor data may lead to novel privacy issues. To address these issues, a user-centric understanding of technology acceptance and concerns is essential. Therefore, to this end, we conducted a large-scale crowdsourcing study with 1036 participants, examining user decision-making processes regarding LLM-powered conversational agents in XR, across factors of XR setting type, speech interaction type, and data processing location. We found that while users generally accept these technologies, they express concerns related to security, privacy, social implications, and trust. Our results suggest that familiarity plays a crucial role, as daily generative AI use is associated with greater acceptance. In contrast, previous ownership of XR devices is linked to less acceptance, possibly due to existing familiarity with the settings. We also found that men report higher acceptance with fewer concerns than women. Regarding data type sensitivity, location data elicited the most significant concern, while body temperature and virtual object states were considered least sensitive. Overall, our study highlights the importance of practitioners effectively communicating their measures to users, who may remain distrustful. We conclude with implications and recommendations for LLM-powered XR. 
\end{abstract}


\section{Introduction}
Extended reality (XR), covering mixed, augmented, and virtual reality (MR/AR/VR) setups, has become ubiquitous with the availability of affordable consumer-grade head-mounted displays (HMDs). These technologies provide numerous benefits in different domains, including training~\cite{Abich2021, lea_and_schlittmeier_2024, acm_sap_2019_b}, maintenance~\cite{Breitkreuz_etal_2022, GUO201841, ALAM20251545}, and entertainment~\cite{weerachet_etal_2022, hock_etal_2017, MULLER2025108751}. Unlike traditional devices, such as laptops or tablets, HMDs offer immersive user experiences and enable the collection of richer data types, including eye tracking~\cite{bozkir2025eyetrackedvirtualrealitycomprehensive} and gestures~\cite{LI201984}. While such data can be helpful for personalized user interfaces, the availability of fine-grained sensor data also raises privacy concerns, as it can be used to derive detailed user profiles. 

An active line of research in digital technologies concerns user companionship~\cite{Chaturvedi_etal_2024, Dodd_etal_2025}. In XR, this has often been approached through embodied conversational agents~\cite{Heerink2010, bozkir_etal_2021}. Cassell defines embodied conversational agent interfaces as replicating human face-to-face conversational abilities, including ``recognizing and responding to verbal input'' and ``giving signals that indicate the state of the conversation, as well as contributing new propositions to the discourse''~\cite{Cassell_2000}. Until recently, conversational characters and avatars in XR have largely relied on pre-scripted or single-purpose agents~\cite{gao_etal_2021, bozkir_etal_2024}, which limits their ability to process varied verbal inputs or sustain coherent conversational states. These shortcomings compromise the authenticity of interactions in XR environments and the ecological validity of applications built on them. With the public release of ChatGPT~\cite{chatgpt_public_2022} and the growing prevalence of large language models (LLMs), it has become feasible to integrate natural, high-quality conversational abilities into XR agent interfaces. As a result, such advances will likely increase user engagement in immersive spaces and lead to privacy-sensitive self-disclosures in a comfortable way~\cite{Zhang_etal_2024, bozkir_etal_2024}. When further combined with sensor data streams, these disclosures can introduce novel privacy risks that may negatively affect users.

Considering the opportunities and challenges, addressing privacy issues in these novel settings is essential to achieve user-centered, immersive XR. To this end, it is first necessary to have a detailed understanding of what users accept and what concerns they hold. Therefore, we ask two research questions (RQs): \textbf{RQ1}: ``To what extent, and under which conditions, do users accept LLM-powered XR technologies?'' and \textbf{RQ2}: ``What kinds of concerns do users raise regarding different conditions and data types?''

To address our RQs, we conducted a large-scale crowdsourcing study ($n=1036$) on LLM-powered XR using a $2\times2\times3$ between-subjects design. We examined three factors: XR setting type (i.e., mixed vs. virtual reality), speech modality type (i.e., basic voice commands vs. generative AI), and data processing location (i.e., local XR device, own server, application cloud of speech service provider). We collected responses on the Unified Theory of Acceptance and Use of Technology 2 (UTAUT2)~\cite{venkatesh2012consumer} and Mixed Reality Concerns (MRC)~\cite{MRC_Q_chi_24} questionnaires, as well as users’ opinions on the sensitivity of different data types and their extraction possibilities. Our analyses reveal two key insights: \textbf{(i)} users generally accept the technology but express privacy and security concerns, alongside trust issues, independent of study factors; and \textbf{(ii)} while most participants recognize the types of data that can be extracted from HMDs, they report the highest concern for location data and the lowest concern for body temperature and the states of virtual objects. As no prior work has systematically investigated these aspects for XR, our study provides a baseline and contributes insights to support the development of responsible solutions for XR. 

\section{Related Work}
Our work focuses on the use of LLM-powered agents in XR and users' perspectives toward conversational agents in these settings. Therefore, we review the prior work in the following sections accordingly.  

\subsection{LLM-powered XR Spaces}
LLM integration into XR has been a focus of many~\cite{delatorre_etal_2024_chi, 10896144}, as it encompasses various opportunities for interactions~\cite{Tang_etal_2025} and LLMs can be potentially helpful for a wide range of tasks. To enable this, Wang et al.~\cite{wang2023chat3d} have focused on combining LLMs and 3D object representations, proposing a data-efficient training scheme for multimodal LLMs that allows the model to learn object attributes and spatial object relations. The proposed method, Chat-3D, demonstrated significant performance in understanding diverse instructions about 3D scenes and in engaging in spatial reasoning. Similarly, Hong et al.~\cite{hong20233d} proposed to inject 3D worlds into LLMs and introduced 3D-LLMs. The authors have demonstrated that by feeding 3D point clouds into LLMs, 3D-LLMs can perform a range of tasks relevant to immersive spaces, including captioning, 3D question answering, and navigation, and they outperform 2D models in 3D captioning and 3D-assisted dialogues. 

More directly related to immersive use-cases, previous work has also focused on creating and editing 3D immersive spaces~\cite{delatorre_etal_2024_chi, 10918869}. While De la Torre et al.~\cite{delatorre_etal_2024_chi} demonstrated positive user experiences with their system, Chen et al.~\cite{10918869} suggested that when they created 3D worlds using JSON data generated by LLMs, there was a significant reduction in task completion times. Similarly, Huang et al.~\cite{huang2024realtimeanimationgenerationcontrol} proposed utilizing LLMs for real-time animation generation and control on rigged 3D models, and they demonstrated that LLMs have a decent potential for flexible state transitions across animations. Furthermore, Guinchi et al.~\cite{10494096} proposed using LLMs to translate natural language into software, thereby democratizing behavioral design for VR. The authors found that while the speech interface is helpful for elementary programming tasks, several challenges need to be addressed, including error handling, incorporating users' feedback, and potential security issues. 

Even though LLMs provide numerous advantages in creating and editing immersive spaces, one of the most straightforward use cases with LLMs is to embed them as conversational agents in XR, as it is possible to utilize them with speech-to-text and text-to-speech models~\cite{bozkir_etal_2024}, as well as in an end-to-end fashion~\cite{realtime_api_openai}. Therefore, many prior works have focused on this direction, exploring user interactions with LLM-powered conversational agents for purposes such as personalization and engagement. To this end, Lau et al.~\cite{lau_etal_eccvw} demonstrated in the context of intangible cultural heritage that users find XR experiences more usable and engaging with LLM-powered chatbots than with pre-scripted chatbots. Furthermore, in another study on cultural heritage, users indicated more engagement and learning interest when LLM-powered personalized conversational agents exist in immersive experiences~\cite{lau2024wrappedanansiswebunweaving}. Similarly, Gao et al.~\cite{10765410} showed the effectiveness and friendliness of learning applications when learners engage with LLM-powered conversational AI agents in VR, in the context of AI literacy training. 

However, despite all the advantages, when such agents can facilitate natural conversational flows, users may self-disclose potentially sensitive information due to high levels of engagement. This sensitive information, combined with behavioral data collected through HMD sensors, may lead to novel privacy invasions for users. Therefore, it is also essential to understand users' perspectives on such technology, particularly in terms of their acceptance and concerns regarding security, privacy, and trust, so that a user-centered approach can be facilitated with responsible data science practices. 

\subsection{Users' Perspectives on Conversational Agents and XR}
Users' perspectives on conversational user interfaces, voice assistants, and AI agents have been studied in the literature~\cite{ling_etal_2021, volkel_etal_2020, denecke_may_2023, volkel_etal_2020_CHI}. For instance, Völkel et al.~\cite{volkel_etal_2020_CHI} studied personality in speech-based conversational agents and indicated that the Big Five model for human personality is not fully applicable for conversational agent personality, meaning that alternative dimensions are needed. Pradhan and Lazar~\cite{pradhan_and_lazar_2021} emphasized that ethics of deception, reinforcement of social stereotypes, and meeting users' needs are essential for the personality and persona of the conversational agents. 

To meet users' needs, it is important to understand their perceptions and acceptance of conversational agents. To this end, Chin et al.~\cite{chin_etal_2024} investigated how formal and information conversational styles affect participants' perceptions and preferences, and they found that older participants preferred informal voice assistants, suggesting a need for personalized conversational designs. Völkel et al.~\cite{volkel_etal_2022_CHI} explored how varying levels of extraversion in chatbots affect users' perception and engagement, and indicated that while participants preferred the extraverted agent, they communicated more with the introverted one, emphasizing the importance of adaptability for user preferences. 

Considering LLMs and their use as conversational agents, Ha et al.~\cite{ha_etal_2024_chi} examined the interactions between users and LLM-based conversational agents. The authors found that customization of these agents facilitates emotional bond and provides sustainable engagement compared to standard LLM interactions. Zhong et al.~\cite{Zhong03022025} explored domain-specific and prompt-driven conversational agents based on LLMs, and they showed that users engage in meaningful multi-turn interactions, which demonstrated LLMs' effectiveness. Beyond perceptions and acceptance, risks and potential harms associated with conversational AI agents are also important, as these agents receive and process data that may include sensitive information from users. To this end, Gumusel et al.~\cite{gumusel2024userprivacyharmsrisks} developed a comprehensive framework to identify and categorize privacy harms and risks in chatbot interactions, providing a guide for responsible design of conversational AI. Despite these tools and implications, embedding such conversational agents into XR presents an opportunity to combine different data sources, such as eye tracking, video streams, and body movements. To fully cover users' perspectives on this novel technology that combines conversational AI and XR, it is also essential to understand users' perspectives on XR. 

When it comes to XR and users' perspectives, a significant amount of prior works focused on privacy~\cite{gallardo2023speculative, bozkir2025eyetrackedvirtualrealitycomprehensive, bozkir2025impactdevicetypedata}. For instance, Gallardo et al.~\cite{gallardo2023speculative} investigated speculative privacy concerns related to AR glasses across various data uses and types. The authors suggested that multidimensional privacy solutions are necessary due to context-dependent concerns. Furthermore, Lee et al.~\cite{lee2016information} explored information disclosure concerns of users towards HMDs, when different entities can receive potentially sensitive data, and found that users become uncomfortable when humans can consume the data compared to computers. Bozkir et al.~\cite{bozkir2025impactdevicetypedata} found similar privacy concerns towards other humans than computers in an AR glasses study, focusing more on eye-tracking data and potential inferences from that with machine learning. Steil et al.~\cite{steil_etal_2019} also investigated attitudes towards eye-tracking data in AR and VR. They implied that users are comfortable with utilizing this data for purposes such as early disease detection or hands-free interaction. 

Lebeck et al.~\cite{lebeck_etal_2018} addressed the security and privacy of multi-user AR, highlighting users' concerns about their physical surroundings. Furthermore, Koelle et al.~\cite{koelle_etal_2017} explored the acceptability and usability of HMDs, finding that social acceptability is more crucial for the short-term adoption of these devices, whereas usability is essential for long-term adoption. Similarly, in another work, Koelle et al.~\cite{koelle_etal_2015} investigated users' attitudes and concerns about HMDs, and the authors identified a few concerning situations, such as recording videos with these devices. However, despite extensive research on HMDs for MR, AR, and VR, it remains an open question how users' technology acceptance and concerns are shaped, especially when novel conversational AI agents, facilitated by generative AI and LLMs, are embedded into XR settings and devices. We address these issues in our work. 

\section{Methods}
We designed a study with several factors to understand users' acceptance and concerns toward LLM-powered conversational agents in XR. The ethics committee of the Technical University of Munich (TUM) approved our study protocol. This section outlines the details of our experimental design, procedure, participants, and data analysis methods. 

\subsection{Experimental Design}
To address our RQs, we utilized a $2\times2\times3$ factorial design in a between-subjects experiment, where participants were allocated to conditions randomly. For independent variables, we used XR setting type, speech interaction type, and speech data processing location. The XR setting type encompasses MR and VR conditions, whereas the speech interaction type involves interaction with specific voice commands and generative AI (i.e., LLMs). Furthermore, the speech processing location encompasses three conditions: on-device (i.e., processing occurs within the head-mounted display), the own server, and the application cloud of the speech interaction service provider. We used these independent variables for the following reasons. Firstly, MR settings overlay virtual content on top of the real content. In principle, one can record the real-world environment, which is considered more privacy-sensitive than solely working in virtual spaces, such as in VR. Secondly, we utilized speech with specific voice commands and LLMs to understand whether a conversational agent that can facilitate naturalistic conversations can increase the acceptance of the technology and affect users' concerns. Lastly, we decided on three conditions for where the speech processing occurs, considering that one can deploy local and lightweight LLMs on local edge devices or their own servers to deal with potential data privacy issues, or use application programming interfaces (APIs) of LLM service providers, where the data processing happens in an application cloud. All these factors may influence users' perspectives regarding their acceptance of the technology and shape their concerns, as well. 

To understand these effects, similar to the recent work~\cite{katins_etal_chi25}, after relevant briefing on the conditions, we designed a survey flow that first includes a UTAUT2 questionnaire~\cite{venkatesh2012consumer} (excluding price value construct due to its irrelevance), followed later by the MRC questionnaire~\cite{MRC_Q_chi_24}. We administered 5-point Likert scales as responses, and the scale of these responses ranged from ``strongly disagree'' to ``strongly agree'' with the middle option of ``neutral.'' Upon completion of these questionnaires, we asked participants to rank their concerns from most concerning to least about particular data types that can be potentially extracted from XR setups, including audio, body temperature, brain waves, eye tracking, heart rate, images, location, movements, reaction times, text, videos, and virtual objects' states, their opinions on whether these data types can be extracted from the mentioned XR setups or not, and whether they could think of any other XR glasses capability that would make them uncomfortable. We provided the response options for concern rankings and extractability perception in a randomized fashion. Lastly, we asked about demographics. We provide full text for our surveys in the Appendix. 

\subsection{Recruitment, Procedure, and Participants} 
To recruit participants for our crowdsourcing study, we used the Prolific platform due to its high-quality data~\cite{Peer_etal_2022}. After participants expressed interest in participating in our study on Prolific, we forwarded them to our survey on Qualtrics, and they first encountered an informed consent form. Once they provided digital consent, they were randomly allocated to an experimental condition and received a briefing and introductory text, along with a descriptive image that corresponded to their allocated condition. They then proceeded with our surveys, and upon completion, we forwarded them back to Prolific for compensation. 

We recruited 1103 participants, who are fluent in English and reside in Germany, with an equal sex distribution option, for a survey that was expected to take 11 minutes. We compensated our participants with \euro{2.75}. To obtain high-quality data, we filtered out participants who reported in our survey that they do not use the Internet despite filling out our survey online, whose reCAPTCHA score is below 0.5 (i.e., indicating a potential bot response on Qualtrics), whose age mismatch between Prolific and Qualtrics is greater than one, and who provided invalid answers to our questions more than six times (consistent with number of distinct pages for our questions). Through this process, we ultimately had 1036 participants ($M_{age}\pm SD_{age} = 31.38\pm 9.35$), including 491 women, 530 men, 11 non-binary, and 4 who preferred not to describe their gender, for analysis. Of our participants, 1004 have used generative AI-based services before, while 854 identified themselves as current users of such services. Furthermore, 190 of our participants had previously owned XR glasses, while 96 identified themselves as current users of XR. 

\subsection{Analysis}
We employed a general linear model for our primary data analysis, considering our three experimental factors: XR setting type, speech interaction type, and data processing location. We treated VR, voice command-based interaction, and on-device data processing as reference conditions. To account for individual differences, we included several covariates, including age, gender, education, daily Internet use, general generative AI use, daily generative AI use, current XR use, and XR device ownership. The UTAUT2 and MRC questionnaires include different constructs with multiple questions. For UTAUT2, these include performance expectancy (UTAUT2\_PE), effort expectancy (UTAUT2\_EE), social influence (UTAUT2\_SI), facilitating conditions (UTAUT2\_FC), hedonic motivation (UTAUT2\_HM), habit (UTAUT2\_HT), and behavioral intention (UTAUT2\_BI), whereas for MRC, these are security \& privacy (MRC\_SP), social implications (MRC\_SI), and trust (MRC\_T). For the UTAUT2, higher Likert responses correspond to a greater acceptance. For MRC, the trust construct is reverse-coded compared to constructs on security \& privacy and social implications. Therefore, higher score responses in MRC\_SP and MRC\_SI correspond to greater concern, whereas lower score responses in MRC\_T correspond to less trust. When calculating dependent variables for all these constructs, we computed the mean values of the Likert scale responses from multiple questions within each construct, and treated these as approximately continuous variables. For model estimation, we utilized ordinary least squares (OLS) regression with HC3 heteroskedasticity-consistent robust standard errors, which provides valid analyses even when residuals deviate from normality or show heteroskedasticity. We utilized a Type II ANOVA based on the fitted regression model to test for the main and interaction effects of our experimental factors, with an alpha level of $0.05$.

In addition to the primary analysis, we examined participants’ concerns regarding different data types and their perceptions of data extraction possibilities in XR settings. For the former, participants ranked multiple data types from most to least concerning. We analyzed these rankings using a linear mixed-effects model with data type as a within-subjects factor and the experimental factors (i.e., XR setting type, speech interaction type, and data processing location) as between-subjects factors. Each participant’s responses were treated as repeated measures, and a random intercept for each participant accounted for individual differences in overall ranking tendencies. The assigned rank value served as the dependent variable. We treated the audio data category as the reference level in this analysis, and all comparisons of data-type concern were made relative to audio, reflecting that communication with conversational agents in XR primarily occurs through audio in our described scenario. Omnibus significance tests for the fixed effects were obtained using Wald $\chi^2$ tests, which provide ANOVA-like inferences for mixed-effects models, with an alpha level of $.05$. For the latter, we analyzed participants’ perceptions of the data extraction possibilities of each data type descriptively in an exploratory manner. 

\section{Results}
\begin{figure}[t!]
 \centering
   \includegraphics[width=\linewidth]{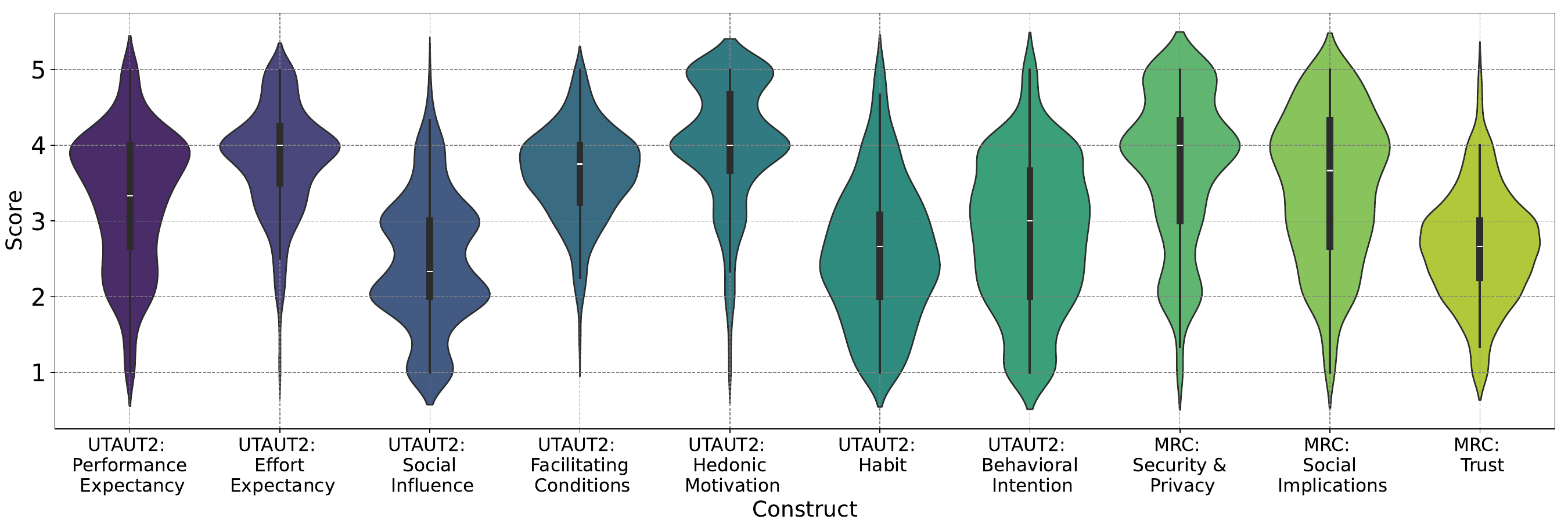}
    \caption{Survey response distributions for the different constructs of UTAUT2 and MRC. The Trust construct in the MRC questionnaire is reverse-coded according to the original survey.}
    \label{fig_aggregated_responses}
\end{figure}

Our analyses on the main experimental factors (i.e., XR setting type, speech interaction type, and data processing location) for each of the constructs in UTAUT2 and MRC questionnaires did not provide enough evidence that participants' technology acceptance and concerns toward conversational agents in XR differ based on whether they interact with these agents in MR or VR, whether the conversations are facilitated through basic voice commands or LLMs, and whether the speech processing happens on their own device, own server, or application cloud (through APIs) ($p > .05$). We depict the aggregated response distributions across conditions for each of the constructs we studied in UTAUT2 and MRC in Figure~\ref{fig_aggregated_responses}. Our results on the main experimental factors particularly indicate that our participants did not have additional concerns towards novel approaches, such as conversational agents through LLMs in XR. However, in all conditions, our participants were generally concerned, regardless of the assigned condition. In addition, similar to concerns, our participants' technology acceptance is also comparable and relatively high in the majority of the constructs in UTAUT2; it is reasonable to state that users are open to using these novel technologies. We provide summary statistics of condition-wise distributions in Table~\ref{sum_stats_conditions}. 

\begin{table}[h!]
\caption{Summary statistics of condition-wise survey responses. The Trust construct in the MRC questionnaire is reverse-coded according to the original survey.}
\centering
\resizebox{\textwidth}{!}{%
\begin{tabular}{c|c|c|c|c|c|c|c|c}
 &  & \multicolumn{7}{c}{\textbf{Experimental Conditions ($M \pm SD$)}} \\ \hline
 & \textbf{Construct} & \textbf{MR} & \textbf{VR} & \textbf{Voice commands} & \textbf{LLMs} & \textbf{On-device} & \textbf{Own server} & \textbf{App cloud} \\ \hline
\multirow{7}{*}{\rotatebox[origin=c]{90}{\textbf{UTAUT2}}} 
 & Performance Expectancy & $3.29 \pm 0.90$ & $3.31 \pm 0.89$ & $3.27 \pm 0.90$ & $3.34 \pm 0.89$ & $3.30 \pm 0.90$ & $3.34 \pm 0.91$ & $3.26 \pm 0.88$ \\ \cline{2-9}
 & Effort Expectancy & $3.76 \pm 0.72$ & $ 3.82\pm 0.70$ & $ 3.82\pm 0.67$ & $3.76 \pm 0.72$ & $3.77 \pm 0.66$ & $3.81 \pm 0.70$ & $3.80 \pm 0.73$ \\ \cline{2-9}
 & Social Influence & $2.41 \pm 0.84$ & $2.51 \pm 0.87$ & $2.44 \pm 0.97$ & $2.48 \pm 0.84$ & $2.42 \pm 0.87$ & $2.53 \pm 0.85$ & $2.42 \pm 0.84$ \\ \cline{2-9}
 & Facilitating Conditions & $3.61 \pm 0.59$ & $3.65 \pm 0.63$ & $3.65 \pm 0.61$ & $3.61 \pm 0.62$ & $3.63 \pm 0.59$ & $3.63 \pm 0.64$ & $3.63 \pm 0.60$ \\ \cline{2-9}
 & Hedonic Motivation & $4.02 \pm 0.81$ & $4.03 \pm 0.82$ & $4.07 \pm 0.81$ & $3.99 \pm 0.81$ & $4.01 \pm 0.82$ & $4.04 \pm 0.82$ & $4.04 \pm 0.79$ \\ \cline{2-9}
 & Habit & $2.56 \pm 0.93$ & $2.58 \pm 0.90$ & $2.58 \pm 0.92$ & $2.56 \pm 0.91$ & $2.58 \pm 0.93$ & $2.63 \pm 0.93$ & $2.50 \pm 0.88$ \\ \cline{2-9}
 & Behavioral Intention & $2.80 \pm 0.98$ & $2.85 \pm 0.97$ & $2.85 \pm 0.98$ & $2.80 \pm 0.97$ & $2.81 \pm 0.99$ & $2.89 \pm 0.99$ & $2.77 \pm 0.96$ \\ \hline
\multirow{3}{*}{\rotatebox[origin=c]{90}{\textbf{MRC}}}
 & Security \& Privacy & $3.71 \pm 0.99$ & $3.67 \pm 1.00$ & $3.68 \pm 0.99$ & $3.70 \pm 1.00$ & $3.64 \pm 1.02$ & $3.67 \pm 1.02$ & $3.76 \pm 0.94$ \\ \cline{2-9}
 & Social Implications & $3.48 \pm 0.94$ & $3.42 \pm 0.99$ & $3.42 \pm 0.97$ & $3.49 \pm 0.96$ & $3.47 \pm 0.97$ & $3.41 \pm 1.00$ & $3.48 \pm 0.92$ \\ \cline{2-9}
 & Trust & $ 2.67\pm 0.71$ & $2.63 \pm 0.75$ & $2.65 \pm 0.73$ & $2.66 \pm 0.74$ & $2.67 \pm 0.74$ & $2.68 \pm 0.71$ & $2.61 \pm 0.74$ \\ \hline
\end{tabular}
}
\label{sum_stats_conditions}
\end{table}

When we examined the additional factors that affect technology acceptance and concerns through our covariates, statistically significant patterns emerged. For UTAUT2, we found that men provided significantly higher ratings than women for all constructs, including habit ($p < .01$), and the remaining constructs ($p < .0001$). Furthermore, individuals who owned XR glasses provided lower scores in different constructs of UTAUT2, including effort expectancy, facilitating conditions, and behavioral intention ($p < .0001$), hedonic motivation and habit ($p < .01$), and social influence ($p < .05$), than those who had not owned XR glasses. Furthermore, more daily generative AI use led to higher technology acceptance, as evidenced in the scores for performance expectancy, social influence, hedonic motivation, and behavioral intention ($p < .01$), and habit ($p < .001$)

For MRC, similar to the results of UTAUT2, we found that men were significantly less concerned than women in security \& privacy and in social implications ($p < .0001$). In addition, increased daily use of generative AI led to greater trust in such services ($p < .05$). Lastly, we also found that participants who previously used generative AI services have more concerns in terms of security \& privacy and social implications ($p < .05$). Despite that, since the majority of our participants reported previous use of these services, this finding should be treated cautiously. 

\begin{figure}[t!]
 \centering
   \includegraphics[width=\linewidth]{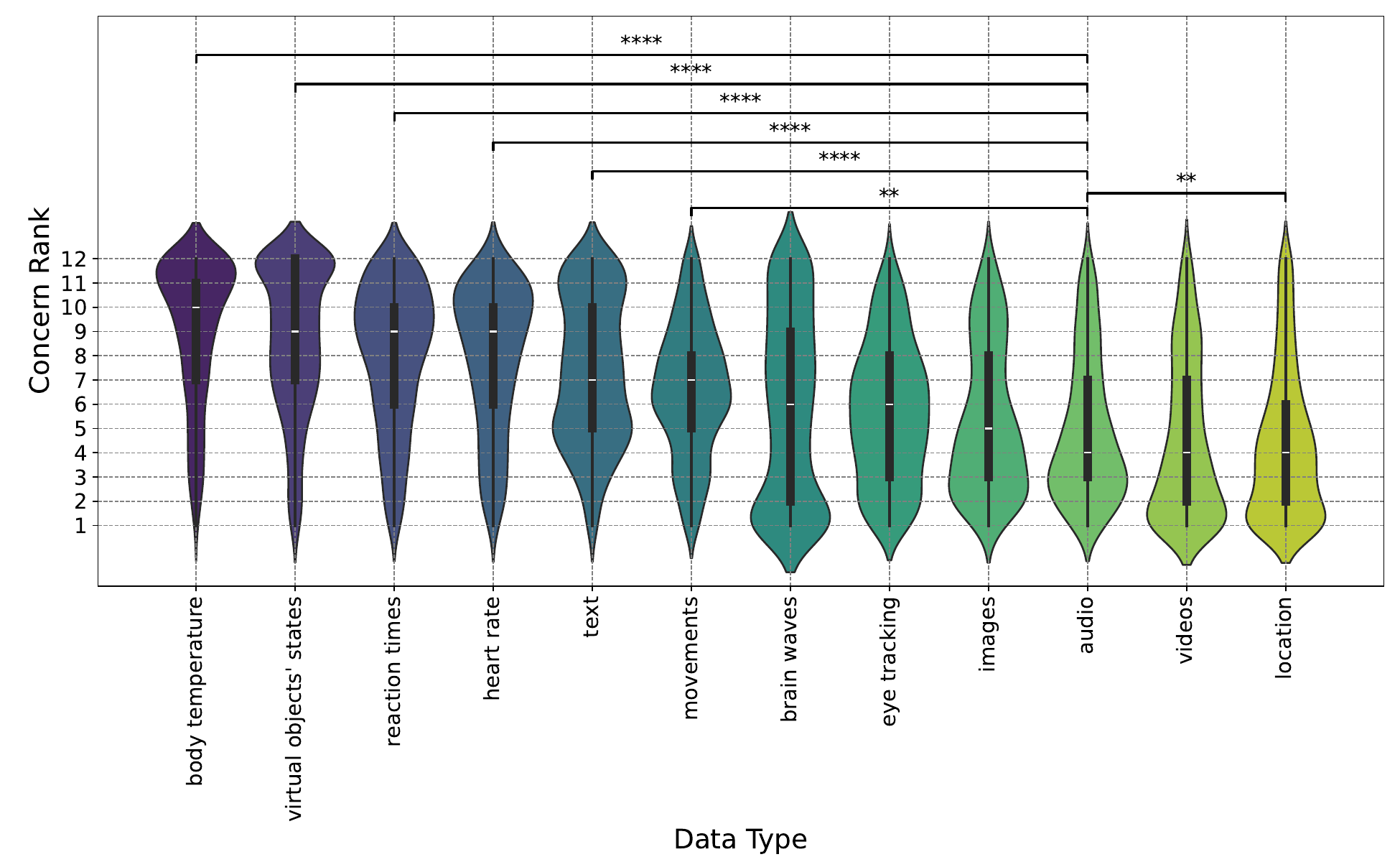}
    \caption{Participants' concerns on different data types. The smaller the number, the higher the concerns participants had. ** and **** correspond to $p < .01$ and $p < .0001$, respectively.}
    \label{fig_data_concern_rank}
\end{figure}

In addition, to understand which data types are the most concerning, we asked our survey participants to rank the data types from most concerning to least. We then compared the concerns by using audio data as a reference level, due to the inherent relationship between audio data and conversational agents. We found that our participants ranked location ($M = 4.34 \pm 3.09$) as the most concerning data type, and its ranking is significantly higher than audio ($M = 4.94 \pm 2.96$) with ($p<.01$). While the concern rankings for audio, videos ($M = 4.5 \pm 3.25$), images ($M = 5.35 \pm 3.07$), eye tracking ($M = 5.67 \pm 2.87$), and brain waves ($M = 5.84 \pm 3.86$) are comparable, participants were significantly less concerned about movements ($M = 6.55 \pm 2.71$) than audio ($p < .01$). Similarly, concerns toward text ($M = 7.33 \pm 3.01$), heart rate ($M = 7.86 \pm 3.01$), reaction times ($M = 7.94 \pm 2.97$), virtual objects' states ($M = 8.63 \pm 3.06$), and body temperature ($M = 9.05 \pm 2.95$) were significantly less than concerns toward audio ($p < .0001$). We provide these results and relationships in Figure~\ref{fig_data_concern_rank}. 

\begin{figure}[t!]
 \centering
   \includegraphics[width=0.95\linewidth]{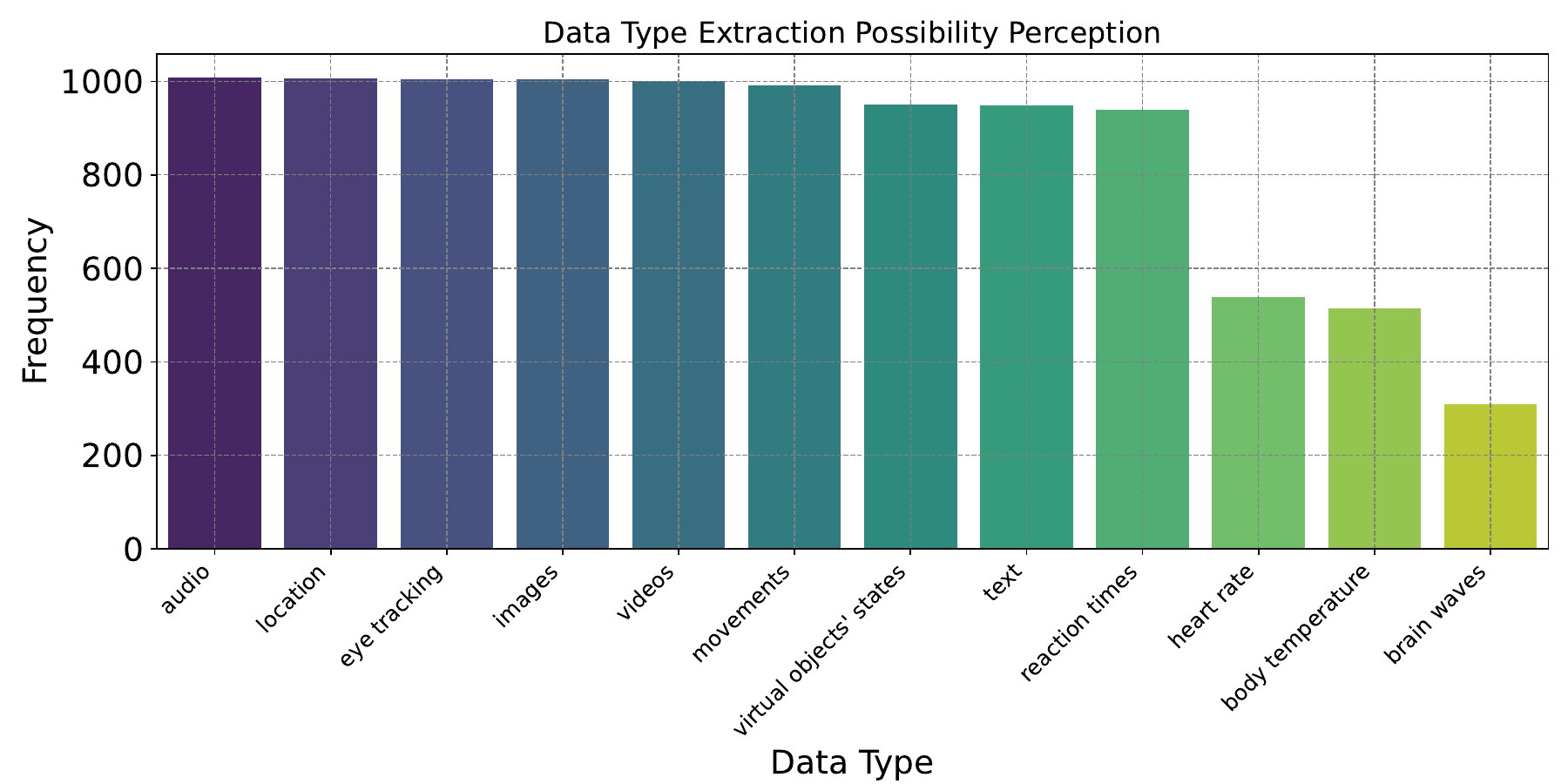}
    \caption{Participants' perception of what type of data can be extracted from the XR settings that include conversational agents. Frequency corresponds to the number of participants who thought the extraction was possible.}
    \label{fig_data_extractability}
\end{figure}

Lastly, to understand how informed our participants were about the settings that included XR and conversational agents, we also analysed their perceptions of the extractability of the data types we presented to them. Most of our participants believed that it is possible to extract data on audio, location, eye tracking, images, videos, movements, virtual objects' states, text, and reaction times from conversational agent-supported XR settings, which reflects the current capabilities of XR HMDs and virtual spaces~\cite{varjo_xr4_2023}. Heart rate, body temperature, and brain waves split our participants' decisions regarding extractability. Overall, it is reasonable to state that our participants understood the settings well in terms of their capabilities and responded to our questions accordingly. We provide this data in Figure~\ref{fig_data_extractability}. 

\section{Discussion}
We discuss our findings in light of our two RQs and provide two recommendations (i.e., \textbf{R1} and \textbf{R2}), in the following. 

\subsection{Insights on User Acceptance}
To address our \textbf{RQ1}, we asked our participants the UTAUT2 questionnaire, considering our experimental design. We did not find sufficient evidence that our participants' opinions differ on whether conversational agents are facilitated in MR or VR. This finding is particularly interesting since users encounter virtual content blended into a real-world environment in MR. In contrast, in VR, users observe fully virtual content, and the potential privacy implications differ significantly. Despite that, we found that our participants accepted such technologies comparably. Technology acceptance towards interacting with conversational agents that are supported by basic voice commands and more sophisticated LLMs did not differ, indicating that our participants accept this type of speech interaction itself, and potentially, more natural speech-based interaction through LLMs did not lead them to additional excitement. Lastly and more interestingly, our participants had comparable levels of acceptance whether speech data processing occurs on-device, on their own server, or on an application cloud, where the relevant data needs to be transferred to a third-party device. Although this process is more closely related to data privacy issues, the potential data transfer did not decrease participants' technology acceptance of conversational AI for XR. We initially hypothesized that when speech interaction is facilitated through an application cloud (i.e., a third-party service), our participants would be less likely to accept such a technology; however, this was not the case according to our analyses. The reason might be that generative AI services, such as ChatGPT~\cite{chatgpt_public_2022} and Gemini~\cite{Deepmind2023}, have democratized access to these services, and potential data privacy issues have not led to a decrease in technology acceptance for our participants. 

Overall, beyond our experimental factors, our participants provided positive feedback for the acceptance of conversational agent-supported XR for performance expectancy, effort expectancy, facilitating conditions, and hedonic motivation constructs. In contrast, they tended to be negative in the social influence construct. For the habit and behavioral intention constructs, they inclined towards being more neutral. \textbf{Considering all, our participants accepted the use of novel conversational AI-supported XR technology.} 

When we examined additional factors beyond the main experimental factors, we found three significant trends. Firstly, men tend to accept this kind of technology more than women across all constructs of UTAUT2. Previous work also suggested that demographics such as age and gender can lead to different attitudes towards XR glasses~\cite{koelle_etal_2015, bozkir2025impactdevicetypedata}. For instance, Koelle et al.~\cite{koelle_etal_2015} reported that women may be less enthusiastic about technology and are more likely to express negative feelings, which aligns with our findings. Furthermore, participants who owned XR glasses showed less acceptance than those who did not in the majority of the constructs. This finding might be due to the novelty of the mentioned technology for the participants who did not own XR devices. Furthermore, XR device owners are likely to be more aware of the capabilities of such devices, which may lead to them being less enthusiastic than others. Lastly, participants who use more generative AI-based services daily expressed higher acceptance toward conversational agent-supported XR, in five of the UTAUT2 constructs. Our findings indicate that the more users are exposed to these novel technologies, especially generative AI, the greater their acceptance of them tends to be. Therefore, as the first recommendation (\textbf{R1}), we suggest implementing public education and training initiatives to increase awareness and familiarity with these technologies. By also encouraging individuals to integrate them into their daily lives and routines, the general public may adopt them more naturally, which would lead to a more informed acceptance of the technology. 

\subsection{Insights on Concerns}
Addressing our \textbf{RQ2}, similar to our findings for UTAUT2, we did not find sufficient evidence in the MRC questionnaire that concerns toward conversational agent-supported XR differ based on our experimental factors. This finding is particularly encouraging, especially for implementing sophisticated conversational AI agents with LLMs, as they do not raise additional concerns compared to previously established, basic voice command-based approaches. In addition, despite MR settings including real-world content (e.g., videos from physical spaces) compared to VR, which is often considered more privacy-sensitive, our participants did not report differing concerns between MR and VR. It is also possible that our participants may have focused more on the conversational AI aspects during our survey, as this direction is relevant in everyday life, given that OpenAI introduced ChatGPT to the general public~\cite{chatgpt_public_2022}, which may have contributed to this finding. Furthermore, as long as the data is processed in a computer (e.g., on-device, own server, or application cloud), participants did not have particularly differing concerns. For AR glasses data collection and processing, Bozkir et al.~\cite{bozkir2025impactdevicetypedata} have recently found that privacy concerns escalate when other humans can access the data, rather than different types of computers, and our findings also confirm this in terms of device comparisons. In addition, our findings also imply comparable comfort between on-device deployment of LLM-based conversational agents and API access, in principle. This finding may also be due to the digital resignation of our participants~\cite{Draper_and_Turow_2019}, which warrants further investigation in the future. \textbf{Overall, we found that our participants did not have additional reservations toward novel, conversational AI-supported XR.} However, regardless of the experimental factor and indifferent concerns toward those factors, our participants have issues about security \& privacy, social implications, and trust, as our scores incline toward a negative direction in MRC. Therefore, (\textbf{R2}) we recommend that, alongside advancing the technological development of these novel technologies, researchers and practitioners should prioritize creating clear, concise, and comprehensible descriptions for everyday users. Providing such comprehensible information may facilitate more comfortable and confident use of these technologies and their associated products, which is essential for addressing diverse user concerns for the future.  

In addition to the primary analyses, when we examine additional factors that affect concerns in MRC, similar to the UTAUT2 results, men indicated significantly less concern in security \& privacy and social implications than women. This finding is also in line with prior work on AR glasses~\cite{bozkir2025impactdevicetypedata}. Furthermore, as the trust construct in the MRC is significantly higher for higher daily use of generative AI, this finding highlights the importance of exposure to novel technologies in increasing the associated trust. 

We also analyzed users' concerns towards different data types. By keeping audio as a reference due to its direct relationship with the communication with conversational agents in XR, we found that the most concerning data types are location, videos, audio, and images. It is interesting that, despite location being tracked in various devices beyond XR HMDs, such as smartphones, tablets, and personal computers, we found location to be the most concerning data type. We initially hypothesized that unusual and medical data types, such as brain waves, body temperature, or heart rate, would gather more concern than the others. However, we did not find such a trend. Considering that the most concerning data types (e.g., location, videos, audio, and images) can be extracted from consumer-grade HMDs, and our participants' understanding of this (Figure~\ref{fig_data_extractability}), this finding is somewhat unforeseen and requires further investigation into the possible reasons for it. 

\subsection{Limitations and Future Work}
We conducted a large-scale crowdsourcing study about conversational agent-supported XR. However, our work has several limitations that require further research. Firstly, almost all our participants had previously used generative AI-based services, and most of them identified themselves as current users. Especially the participants who were assigned to the basic voice-based command condition for the speech interaction type experimental factor might have assumed that interactions are facilitated with LLM-based intelligent agents anyway (due to the ubiquity of LLM-based agents), which may have shaped their responses accordingly. Secondly, due to the level of technology literacy among our participant pool, which includes individuals who use generative AI daily, their perceptions of conversational agent-supported XR may differ from those of the general public. Furthermore, although we have a handful of participants who own XR devices, the majority do not own any device and do not use these devices daily. Actual preferences for these devices are most likely shaped through continuous use, and these should be assessed essentially in a longitudinal manner. Lastly, while we believe that we provide a solid baseline for understanding acceptance and concerns regarding this novel technology intersection between LLMs and XR, our analysis and findings should be supported by extensive qualitative data to understand participants' rationales for providing these responses. Future research should focus on these limitations and extend our findings. 

\section{Conclusion}
In this work, we employed Unified Theory of Acceptance and Use of Technology 2 and Mixed Reality Concerns questionnaires in a crowdsourcing study to assess user acceptance and concerns about conversational agent-supported XR through LLMs. Our user study utilized a factorial design, including the XR setting type (i.e., MR and VR), speech interaction type (i.e., basic voice commands and LLMs), and data processing location (i.e., on-device, own server, and application cloud). We did not find sufficient evidence to suggest that user acceptance and concerns differ based on these factors. In addition, we found that participants generally accepted this novel technology, with concerns about security \& privacy and social implications. Furthermore, when we examined additional factors, we found that participants were most concerned about location data and least concerned about body temperature and the states of virtual objects. We provided implications and recommendations for researchers and practitioners. 

\section*{Acknowledgments}
The authors used GPT-5 and Grammarly to improve their paper. They carefully reviewed all the content and updated it as necessary. This work was supported by the Friedrich Schiedel Fellowship. The authors acknowledge the financial support provided by the fellowship, which was instrumental in enabling this research. The fellowship is hosted by the TUM School of Social Sciences and Technology and the TUM Think Tank at the Munich School of Politics and Public Policy.

\bibliographystyle{unsrt}   % or abbrv, unsrt, alpha, etc.
\bibliography{references}   % references.bib is your BibTeX file

\newpage

\appendix
\section{Appendix A}
In this section, we provide details of the user study as follows. 

\subsection{Recruitment}
We used the following recruitment text on Prolific.

``Study participants wanted!

Researchers from the Technical University of Munich (TUM) and TUM Think Tank at the Munich School of Politics and Public Policy (HfP) are looking for participants for an 11-minute survey study on extended reality glasses covering the broad spectrum of virtual and mixed reality. You will be answering several questions regarding the use of extended reality glasses. You will receive $\approx$ 2.75 EUR for your participation.

Individuals aged 18 or older, who are located in Germany, can participate in this study.''

\subsection{Briefing}
Participants encountered the following briefing texts based on the assigned experimental condition. Parts of the briefing text were inspired by the prior work~\cite{bozkir2025impactdevicetypedata}.

``In the following pages, you are expected to answer some questions about [XR type] glasses. You do not necessarily need to have any experience with [XR type] glasses.''

\textbf{Mixed reality:} ``Imagine that you are an owner of mixed reality glasses that are worn on your head and can perceive your surroundings and your own state with the help of its sensors. In addition, your mixed reality glasses can display virtual objects in your field of view, and these objects can benefit you during different tasks interacting with the real world. To facilitate such functionality, your mixed reality glasses can record your surroundings, where you look, what you say, and your overall circumstances in a non-intrusive way. Similar to other smart devices like smartphones, you can install applications in your mixed reality glasses that can help you in your daily activities.''

\textbf{Virtual reality:} ``Imagine that you are an owner of virtual reality glasses that are worn on your head and can perceive your surroundings and your own state with the help of its sensors. In addition, your virtual reality glasses display fully immersive virtual spaces in your field of view, and these virtual spaces can benefit you during different tasks. To facilitate such functionality, your virtual reality glasses can record the virtual scenes you observe, where you look, what you say, and your overall circumstances in a non-intrusive way. Similar to other smart devices like smartphones, you can install applications in your virtual reality glasses that can help you in your daily activities.''

\textbf{Speech interaction with basic voice commands:} ``During the use of your [XR type] glasses, you have the opportunity to interact with a speech-based system that can understand specific voice commands and support you.''

\textbf{Speech interaction with generative AI:} ``During the use of your [XR type] glasses, you have the opportunity to interact with a speech-based chatbot that is powered by generative artificial intelligence (e.g., ChatGPT, Copilot, Claude). Such chatbots can accurately understand what you tell them, communicate with you naturally, and support you.''

\textbf{Data processing on-device:} ``Suppose that data processing for the speech interaction happens locally within your [XR type] glasses.''

\textbf{Data processing on own server:} ``Suppose that data processing for the speech interaction happens on your own server.''

\textbf{Data processing on application cloud:} ``Suppose that data processing for the speech interaction happens on the application cloud of the speech interaction service provider.''

\subsection{Survey Details}
In this section, we provide the detailed text used for each questionnaire, as follows. For the UTAUT2~\cite{venkatesh2012consumer} and MRC~\cite{MRC_Q_chi_24} questionnaires, we used 5-point Likert scales for responses, ranging from ``strongly disagree'' to ``strongly agree'' with the middle option of ``neutral.'' 

\subsubsection{Unified Theory of Acceptance and Use of Technology 2 (UTAUT2)}
\begin{itemize}
  \item UTAUT2\_PE1: I would find these [XR type] glasses useful in my daily life.
  \item UTAUT2\_PE3: Using these [XR type] glasses would help me accomplish things more quickly.
  \item UTAUT2\_PE4: Using these [XR type] glasses would increase my productivity.
  \item UTAUT2\_EE1: Learning how to use these [XR type] glasses would be easy for me.
  \item UTAUT2\_EE2: My interaction with these [XR type] glasses would be clear and understandable.
  \item UTAUT2\_EE3: I would find these [XR type] glasses easy to use.
  \item UTAUT2\_EE4: It would be easy for me to become skillful at using these [XR type] glasses.
  \item UTAUT2\_SI1: People who are important to me would think that I should use these [XR type] glasses.
  \item UTAUT2\_SI2: People who influence my behavior would think that I should use these [XR type] glasses.
  \item UTAUT2\_SI3: People whose opinions that I value would prefer that I use these [XR type] glasses.
  \item UTAUT2\_FC1: I would have the resources necessary to use these [XR type] glasses.
  \item UTAUT2\_FC2: I would have the knowledge necessary to use these [XR type] glasses.
  \item UTAUT2\_FC3: These [XR type] glasses would be compatible with other technologies I use.
  \item UTAUT2\_FC4: I could get help from others when I have difficulties using these [XR type] glasses.
  \item UTAUT2\_HM1: Using these [XR type] glasses would be fun.
  \item UTAUT2\_HM2: Using these [XR type] glasses would be enjoyable.
  \item UTAUT2\_HM3: Using these [XR type] glasses would be very entertaining.
  \item UTAUT2\_HT1: The use of these [XR type] glasses would become a habit for me.
  \item UTAUT2\_HT2: I would be addicted to using these [XR type] glasses.
  \item UTAUT2\_HT3: I must use these [XR type] glasses.
  \item UTAUT2\_BI1: I intend to use these [XR type] glasses in the future.
  \item UTAUT2\_BI2: I would always try to use these [XR type] glasses in my daily life.
  \item UTAUT2\_BI3: I plan to use these [XR type] glasses frequently. 
\end{itemize}

\subsubsection{Mixed Reality Concerns (MRC)}
\begin{itemize}
\item MRC\_SP1: I am concerned about the possibility of non-authenticated individuals gaining access to this [XR type] system.
\item MRC\_SP2: I am concerned about the potential exposure of sensitive data through this [XR type] system to unauthorized parties.
\item MRC\_SP3: I worry that using this [XR type] system might lead to my personal information being misused.
\item MRC\_SI1: I fear that with this [XR type] system, it becomes increasingly hard to maintain a clear distinction between virtual behavior and real-life behavior.
\item MRC\_SI2: I am concerned about the potential of this [XR type] system to influence my behaviors in ways that could be detrimental to my well-being.
\item MRC\_SI3: Using this [XR type] system might make me appear disconnected from others in my physical environment.
\item MRC\_T1: I believe that only legitimate individuals can access this [XR type] system.
\item MRC\_T2: I am sure that this [XR type] system is maintaining a secure environment.
\item MRC\_T3: I am confident that my anonymity is protected by this [XR type] system.
\end{itemize}

\subsubsection{Rankings, Extractability, and Free-text Concerns}
In the following, questions on the rankings for concerns about data types, extraction possibility, and a free-text question about additional concerns are provided. The answer option orders of the first two questions are randomized to avoid any ordering effect.

\begin{itemize}
\item Please rank the following data types associated with [XR type] glasses from most concerning to least concerning by dragging and dropping the corresponding boxes. 1 indicates the most concerning data type, while 12 indicates the least concerning. For answers, see Table~\ref{table_concern_rank}.

\item In your opinion, which of the following data types can be extracted using [XR type] glasses that are mentioned in this study? For answers, see Table \ref{table_extraction_possibilities}. 

\item What other capabilities of [XR type] glasses would make you uncomfortable? 
\underline{\hspace{5cm}}
\end{itemize}

\begin{table}[h!]
\centering

\begin{minipage}{0.49\textwidth}
\centering
\caption{Concern rank question structure for different data types.}
\begin{tabular}{ll}
\toprule
\textbf{Data type} & \textbf{Concern rank} \\
\midrule
Audio & \underline{\hspace{0.5cm}} \\
Body temperature & \underline{\hspace{0.5cm}} \\
Brain waves & \underline{\hspace{0.5cm}} \\
Eye tracking & \underline{\hspace{0.5cm}} \\
Heart rate & \underline{\hspace{0.5cm}} \\
Images & \underline{\hspace{0.5cm}} \\
Location & \underline{\hspace{0.5cm}} \\
Movements & \underline{\hspace{0.5cm}} \\
Reaction times & \underline{\hspace{0.5cm}} \\
Text & \underline{\hspace{0.5cm}} \\
Videos & \underline{\hspace{0.5cm}} \\
Virtual objects' states & \underline{\hspace{0.5cm}} \\
\bottomrule
\end{tabular}
\label{table_concern_rank}
\end{minipage}
\hfill
\begin{minipage}{0.49\textwidth}
\centering
\caption{Question structure for data extraction possibilities.}
\begin{tabular}{ll}
\toprule
\textbf{Data Type} & \textbf{Possible Not Possible} \\
\midrule
Audio & $\square \qquad \square$ \\
Body temperature & $\square \qquad \square$ \\
Brain waves & $\square \qquad \square$ \\
Eye tracking & $\square \qquad \square$ \\
Heart rate & $\square \qquad \square$ \\
Images & $\square \qquad \square$ \\
Location & $\square \qquad \square$ \\
Movements & $\square \qquad \square$ \\
Reaction times & $\square \qquad \square$ \\
Text & $\square \qquad \square$ \\
Videos & $\square \qquad \square$ \\
Virtual objects' states & $\square \qquad \square$ \\
\bottomrule
\end{tabular}
\label{table_extraction_possibilities}
\end{minipage}

\end{table}

\subsubsection{Demographics}
In the following, we present demographic questions that we have adapted from previous work~\cite{leon_etal_2013, bozkir2025impactdevicetypedata}. \\
``In this part of our study, we will ask you about your demographic information.''

\begin{itemize}
  \item How old are you? (In years) \underline{\hspace{0.5cm}}
  \item What is your gender? $\square$ Woman $\square$ Man $\square$ Non-binary $\square$ Prefer not to say $\square$ Prefer to self-describe as \underline{\hspace{0.5cm}}
  \item Which of the following describes best your primary occupation? $\square$ Administrative support (e.g., secretary, assistant) 
  $\square$ Art, writing, or journalism (e.g., author, reporter, sculptor) $\square$ Business, management, or financial (e.g., manager, accountant, banker) 
  $\square$ Computer engineer or IT professional (e.g., systems administrator, programmer, IT consultant) 
  $\square$ Education (e.g., teacher) 
  $\square$ Engineer in other fields (e.g., civil engineer, bio-engineer) $\square$ Homemaker 
  $\square$ Legal (e.g., lawyer, law clerk) 
  $\square$ Medical (e.g., doctor, nurse, dentist) 
  $\square$ Retired 
  $\square$ Scientist (e.g., researcher, professor) 
  $\square$ Service (e.g., retail clerks, server) 
  $\square$ Skilled labor (e.g., electrician, plumber, carpenter) 
  $\square$ Student 
  $\square$ Unemployed 
  $\square$ Decline to answer 
  $\square$ Other (Please specify): \underline{\hspace{0.5cm}}
  
  \item Which of the following best describes your highest achieved education level? 
  $\square$ Without a general school qualification
  $\square$ Lower secondary school qualification (Haupt-(Volks-)schulabschluss)
  $\square$ Intermediate school qualification (Mittlerer Abschluss)
  $\square$ University entrance qualification (Fachhochschul- oder Hochschulreife)
  $\square$ Completed vocational training (Abgeschlossene Berufsausbildung)
  $\square$ University of Applied Sciences degree (Fachhochschulabschluss)
  $\square$ University degree - Bachelor's/4-year degree (Hochschulabschluss - Bachelor/4-Jahres-Abschluss)
  $\square$ University degree - Master's degree (Hochschulabschluss – Master)
  $\square$ Doctorate/PhD
  
  \item Have you ever owned XR glasses? 
  $\square$ Yes 
  $\square$ No
  \item If you currently use XR glasses, approximately how many hours do you spend wearing XR glasses each day? 
  $\square$ \underline{\hspace{0.5cm}} 
  $\square$ Not applicable
  \item Approximately how many hours do you spend on the Internet each day? $\square$ \underline{\hspace{0.5cm}} 
  $\square$ Not applicable
  \item Have you ever used any generative artificial intelligence-based services (e.g., ChatGPT, Copilot, Claude, etc.)? 
  $\square$ Yes 
  $\square$ No
  \item If you currently use any generative artificial intelligence-based services, approximately how many hours do you spend using them? 
  $\square$ \underline{\hspace{0.5cm}} 
  $\square$ Not applicable
\end{itemize}
\end{document}